\documentclass[runningheads]{llncs}
\usepackage[T1]{fontenc}
\usepackage{graphicx}
\usepackage{hyperref}
\usepackage{amsmath}
\usepackage{color}

\newcommand{\mref}[1]{(\ref{#1})}
\newcommand{\sub}[1]{\textsubscript{#1}}

\newcommand{\E}{\operatorname{E}}
\newcommand{\VAR}{\operatorname{VAR}}
\newcommand{\COV}{\operatorname{COV}}

\newcommand{\Prob}{\operatorname{P}}
\newcommand{\given}{\operatorname{|}}

\newcommand{\X}{\mathbf{X}}
\newcommand{\G}{\mathcal{G}}
\newcommand{\PX}[1]{\Pi_{X_{#1}}}

\newcommand{\D}{\mathcal{D}}
\newcommand{\f}[1]{f_{#1}}
\newcommand{\bb}[1]{\boldsymbol{\beta}_{#1}}
\newcommand{\err}[1]{\varepsilon_{#1}}
\renewcommand{\S}[1]{\mathbf{\Sigma}_{#1}}
\newcommand{\xxi}[1]{\xi_{#1}}
\newcommand{\w}[1]{\mathbf{w}_{#1}}
\newcommand{\M}{\mathcal{M}}
\renewcommand{\L}{\mathbf{L}}
\newcommand{\BF}{\mathit{BF}}

\begin{document}

\title{Causal Networks of Infodemiological Data: \linebreak
       Modelling Dermatitis}
\titlerunning{Causal Networks of Infodemiological Data}

\author{Marco Scutari\inst{1} \and Samir Salah\inst{2} \and
  Delphine Kerob\inst{2,3} \and Jean Krutmann\inst{4,5}}

\authorrunning{M. Scutari et al.}

\institute{
  Istituto Dalle Molle di Studi sull'Intelligenza Artificiale (IDSIA), Lugano,
  Switzerland; \email{scutari@bnlearn.com}
\and
  La Roche-Posay Dermatological Laboratories, Levallois-Perret, France \\
\and
  Department of Dermatology, AP-HP Saint-Louis Hospital, Paris, France \\
\and
  Leibniz Research Institute for
  Environmental Medicine, D{\"u}sseldorf, Germany
\and
  Medical Faculty, Heinrich Heine University, D{\"u}sseldorf, Germany
}

\maketitle

\begin{abstract}

  % 150--250 words.

  Environmental and mental conditions are known risk factors for dermatitis and
  symptoms of skin inflammation, but their interplay is difficult to quantify;
  epidemiological studies rarely include both, along with possible confounding
  factors. Infodemiology leverages large online data sets to address this
  issue, but fusing them produces strong patterns of spatial and temporal
  correlation, missingness, and heterogeneity.

  In this paper, we design a causal network that correctly models these complex
  structures in large-scale infodemiological data from Google, EPA, NOAA and US
  Census (434 US counties, 134 weeks). Our model successfully captures known
  causal relationships between weather patterns, pollutants, mental conditions,
  and dermatitis. Key findings reveal that anxiety accounts for 57.4\% of
  explained variance in dermatitis, followed by NO\sub{2} (33.9\%), while
  environmental factors show significant mediation effects through mental
  conditions. The model predicts that reducing PM\sub{2.5} emissions by 25\%
  could decrease dermatitis prevalence by 18\%. Through statistical validation
  and causal inference, we provide unprecedented insights into the complex
  interplay between environmental and mental health factors affecting
  dermatitis, offering valuable guidance for public health policies and
  environmental regulations.


\end{abstract}

\section{Introduction}
\label{sec:intro}

Causal discovery \cite{zanga22} builds on Bayesian network structure learning,
using additional assumptions to ensure that the \emph{causal networks} (CNs) we
learn capture the true data-generating process instead of simple probabilistic
associations. CNs are of core practical importance in epidemiological modelling
\cite{krieger} because many everyday tasks in clinical practice require us to
answer fundamentally causal questions, and CNs are the most effective tool to do
that.

Established algorithms such as LiNGAM \cite{lingam} have theoretical correctness
guarantees but assume that observations are complete, independent and
identically distributed to score competing models from data. Violating these
assumptions will \emph{bias} causal discovery: algorithms will score candidate
CNs incorrectly and select CNs that misrepresent the data-generating process.

However, these assumptions are too restrictive for epidemiology, where data are
typically incomplete and have spatial, temporal and hierarchical structures.
Studying the interplay between mental and dermatological conditions, their
comorbidities and environmental factors further increases data complexity,
making it unfeasible to design a study that measures all of them simultaneously.
Thus, we resort to \emph{infodemiology} \cite{eysenbach}: we integrate or
substitute epidemiological data with information available from internet
big-data databases. Such data are more heterogeneous than typical
epidemiological data because they have been collected and pre-processed
independently at different times and for other purposes.

This paper details a simple yet powerful causal discovery approach that
correctly models complex infodemiological data to produce a CN of dermatitis,
related mental conditions and environmental factors while accounting for
confounding from socio-demographic factors. We improve on our previous work
\cite{loreal22} by explicitly modelling the complex spatio-temporal structure of
these data, validating the learned CN through expert and statistical
validations, and answering complex epidemiological questions not previously
studied in the literature using causal inference.

\section{Background}
\label{sec:background}

\subsection{Infodemiology for Dermatology and Psychiatry}

Dermatological and mental conditions are commonly investigated in isolation;
their interplay is often ignored, limiting our understanding of their aetiology.
Skin and brain interact in many ways: altered skin barrier function is common to
all inflammatory skin diseases; mental disorders overlap in signs and symptoms;
skin diseases impact mental health, mainly anxiety, depression and attention
deficit hyperactivity disorder; stress, anxiety and depression can aggravate or
precipitate the onset of most inflammatory skin diseases. A preliminary analysis
in our previous work \cite{loreal22} shows their \emph{complex network of
interactions}.

At the same time, both classes of conditions have important environmental risk
factors. Outdoor air pollutants are associated with dermatitis \cite{park}, as
are large temperature variations \cite{engebretsen}, and with broad classes of
mental conditions such as depressive disorders \cite{buoli}. These risk factors
and conditions are also associated with the socio-demographic characteristics of
different populations, and therefore \emph{vary across space and time}.

Infodemiology \cite{eysenbach} allows us to construct epidemiological models of
complex interactions in large populations by drawing on large internet
databases. Designing a longitudinal trial to investigate the aetiology of
\emph{dermatitis} and the interplay of its risk factors across the US for
several years would be impractical. Instead, we will use the records of the
pollution monitoring station from the US Environmental Protection Agency (EPA),
the weather data from the National Oceanic and Atmospheric Administration (NOAA)
and web search records for these conditions from Google (used by $87.2\%$ of
Americans). Medical informatics research has found a high correlation between
the occurrence of search queries and the incidence of the corresponding
diseases, allowing us to use the former as a proxy for the latter. Fusing the
data from these sources enables us to construct a population-level longitudinal
data set, which we describe in Section~\ref{sec:data}.

\subsection{Bayesian and Causal Networks}

Bayesian Networks (BNs) \cite{koller} are graphical models defined over a set of
random variables $\X = \{ X_i, i = 1, \ldots, N \}$ that are associated with the
nodes of a directed acyclic graph (DAG) $\G$. Arcs encode the direct
probabilistic dependencies of each $X_i$ on its parents $\PX{i}$, with graphical
separation implying conditional independence in probability. CNs \cite{zanga22}
augment BNs by making the causal edge assumption: each $X_i$ is completely
determined by a function $\f{i}(\PX{i})$ modelling its data generating
mechanism, with noise originating from separate exogenous variables. We can then
interpret arcs as direct cause-effect relationships.

Learning a BN involves learning its structure and parameters from a data set $\D$.
Structure learning consists in finding the DAG $\G$ that encodes the dependence
structure of $\X$ by maximising $\Prob(\G \given \D)$ or another goodness-of-fit
score; LiNGAM \cite{lingam} is a notable example. Causal discovery further
assumes causal sufficiency (the lack of hidden confounders) to ensure we can
recover the data generating mechanisms correctly. We can then learn each $X_i$'s
parameters from $\D$ given $\G$.

Causal discovery typically assumes that $\D$ comprises independent, identically
distributed samples from the data generating mechanism. Relaxations include
extending CNs with missingness graphs and dynamic BNs, duplicating nodes over
different time points to model autocorrelation \cite{neerlandica19}. Accounting
for spatial dependencies and group heterogeneity also relies on duplicating
nodes across locations, using hierarchical priors \cite{krapu} or variational
approximations \cite{ijar21-laura}. However, modelling state-space data in the
same way is computationally prohibitive due to the resulting explosion in the
number of nodes and parameters \cite{tchetgen}.

\section{Materials and Methods}
\label{sec:methods}

The code for the analysis is available at
\url{www.bnlearn.com/research/aime25} .

\subsection{Infodemiological Data Fusion}
\label{sec:data}

\begin{figure}[t]
  \includegraphics[width=\textwidth]{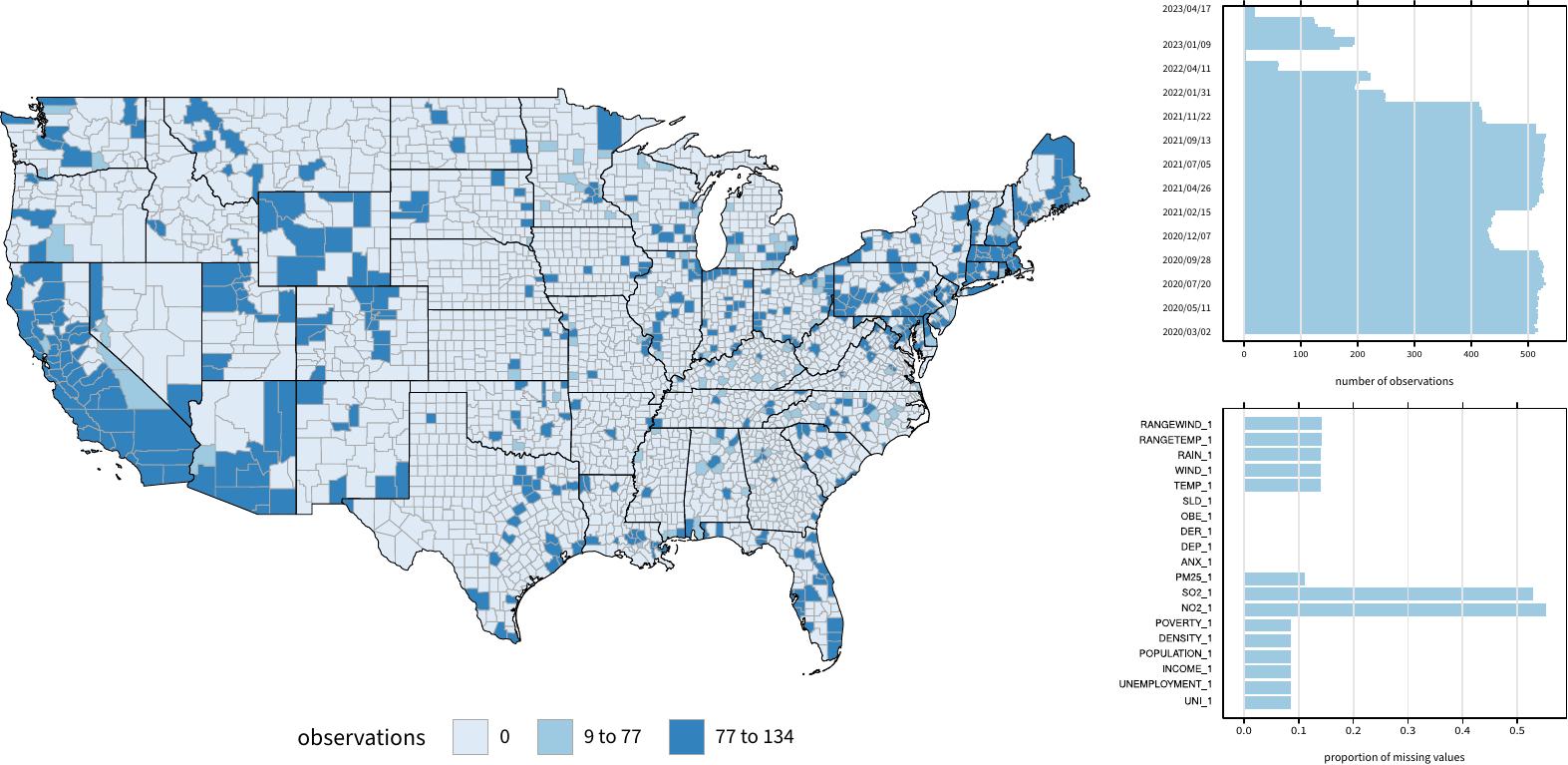}
  \caption{Spatial distribution (left), temporal distribution (top right) and
    proportions of missing values (bottom right) of the infodemiological data.}
  \label{fig:overview}
\end{figure}

We collected 53538 observations over $|\L| = 434$ US counties and 134 weeks from
Google's public data sets \cite{google} and NOAA's Climate Data Online
\cite{noaa}. From Google, we collected weekly measurements of three pollutants
(NO\sub{2}, SO\sub{2}, PM\sub{2.5}, suggested in \cite{park}) from EPA's
``Historical Air Quality'' data three mental conditions (anxiety, depression,
sleep disorders) along with obesity and dermatitis web search frequencies from
Google's ``COVID-19 Public Data Sets.'' NO\sub{2} can also be used as a proxy
for ultrafine (nanosized) particles \cite{no2}. From NOAA, we aggregated weekly
mean and spread measurements for temperature, wind speed and precipitation (as
a proxy for humidity) as suggested in \cite{engebretsen}. To reduce potential
confounding from differences in population and lifestyle, we also included five
socio-demographic variables (the proportion of university graduates,
unemployment rate, poverty rate, household income, population and population
density) from Google's ``US Census Data.'' We treated them as static variables
since their value did not change over the period we covered. Finally, we encoded
each county's centroid latitude and longitude in $\L$.

We fused these data sets using the FIPS county codes and week start dates as
keys. We kept the combination of periods and counties for which each variable
contains at most about 50\% missing values in the fused data set
(Figure~\ref{fig:overview}). The resulting data has strong patterns of missing
values, spatial correlation, temporal correlation and heterogeneity, which we
want to model correctly to elucidate the causal relationships impacting
dermatitis without bias.

\subsection{Model Specification}
\label{sec:definition}

We model the data from Section~\ref{sec:data} with a CN whose nodes are defined
as:
\begin{align}
  &X_{it} = \f{i} (\PX{it}\bb{it}) + \err{it};&
  &\E(\err{it}) = 0, \COV(\err{it}) = \w{it}^\mathrm{T}\S{i}(\L; \xxi{i}) \w{it}.
\label{eq:girls3}
\end{align}
This CN accounts for the complex structure of the data as follows:
\begin{itemize}
  \item \emph{Temporal dependencies} are modelled by duplicating the $X_i$s
    across time points, denoted $X_{it}$. Its parents $\PX{it}$ comprise only
    nodes at time $t - 1$, similar to a dynamic BN.
  \item \emph{Spatial dependencies} are modelled with the covariance matrix
    $\S{i}(\L; \xxi{i})$ as a time-invariant function of geographical location
    $\L$ with parameters $\xxi{i}$ controlling how correlation decays as a
    function of distance.
  \item \emph{Group heterogeneity} is modelled with the weights $\w{it}$.
  \item \emph{Missing data} are accounted for by allowing $\w{it} > 0$ only for
    locally complete (for $X_{it}, \PX{it}, \L\}$) observations, which amounts
    to using the penalised node-averaged likelihood score (PNAL) \cite{ijar21}
    to approximate the marginal likelihood of the CN. We can then compute Bayes
    factors (BFs) as the PNAL ratios.
\end{itemize}

We can estimate $\{ \bb{it}, \w{it}, \xxi{i} \}$ for each $X_{it}$ using a
combination of generalised (GLS) and iteratively reweighted least squares
(IRLS), which provide a general optimisation framework for complex non-linear
dynamics \cite{optim}. Even in the reduced form we consider here, they have much
to offer, as we will see below and in Section~\ref{sec:evaluation}. We do not
consider Expectation Maximisation because its convergence can be very slow
\cite{neerlandica19}, unlike IRLS.

Crucially, \mref{eq:girls3} implies that causal arc directions in the CN are
\emph{completely identifiable} from $\D$. Arcs between nodes in different times
point forward in time. The direction of arcs within the same time point is
always identifiable as well: $X_{it}$ is a non-linear function of $\{\PX{it},
\L\}$ through $\S{i}(\cdot)$ and the $\err{it}$ are heteroscedastic, satisfying
the identifiability conditions from \cite{he2}.

Furthermore, \mref{eq:girls3} makes it easy to \emph{correctly identify
misspecified CNs} by testing the residuals $\err{it}$:
\begin{itemize}
  \item \emph{Temporal dependencies:} test autocorrelation at different lags in
    each location.
  \item \emph{Spatial dependencies:} use Moran's I \cite{moran} at each time
    point, and fit variograms to explore the proportion of variance attributable
    to spatial structure \cite{pinheiro}.
  \item \emph{Heterogeneity:} use Bartlett's test \cite{bartlett} on the
    decorrelated residuals $\S{i}^{-1/2}\err{it}$.
\end{itemize}
We can then adjust the resulting (correlated) p-values for multiplicity using
\cite{benjamini} and compute the proportion of p-values that are greater than a
suitable threshold as a measure of misspecification. In addition, we can assess
the CN's predictive accuracy over unseen times and locations by splitting $\D$
into training and validation subsamples with a sufficiently large buffer in
between \cite{mahoney}.

Finally, \mref{eq:girls3} defines CNs that allow for \emph{probabilistic and
causal inference (queries) spanning both time and space}. We can use them to
compute the posterior probability of events or the distribution of subsets of
$\X$ under specific conditions or after affecting change to the CN \cite{koller}.
Probabilistic queries spanning multiple locations are possible if we augment the
CN with nodes for the $\L$ variables and connect them with all the $X_{it}$. As
for causal inference, interventions over specific locations involve fixing the
corresponding $X_{it}$, while \mref{eq:girls3} holds for the rest; and we
can use the $\sigma$-calculus from \cite{bareinboim} for counterfactuals.
Queries that (also) involve time are constructed similarly to those for dynamic
BNs.

\begin{figure}[h!]
  \includegraphics[width=\textwidth]{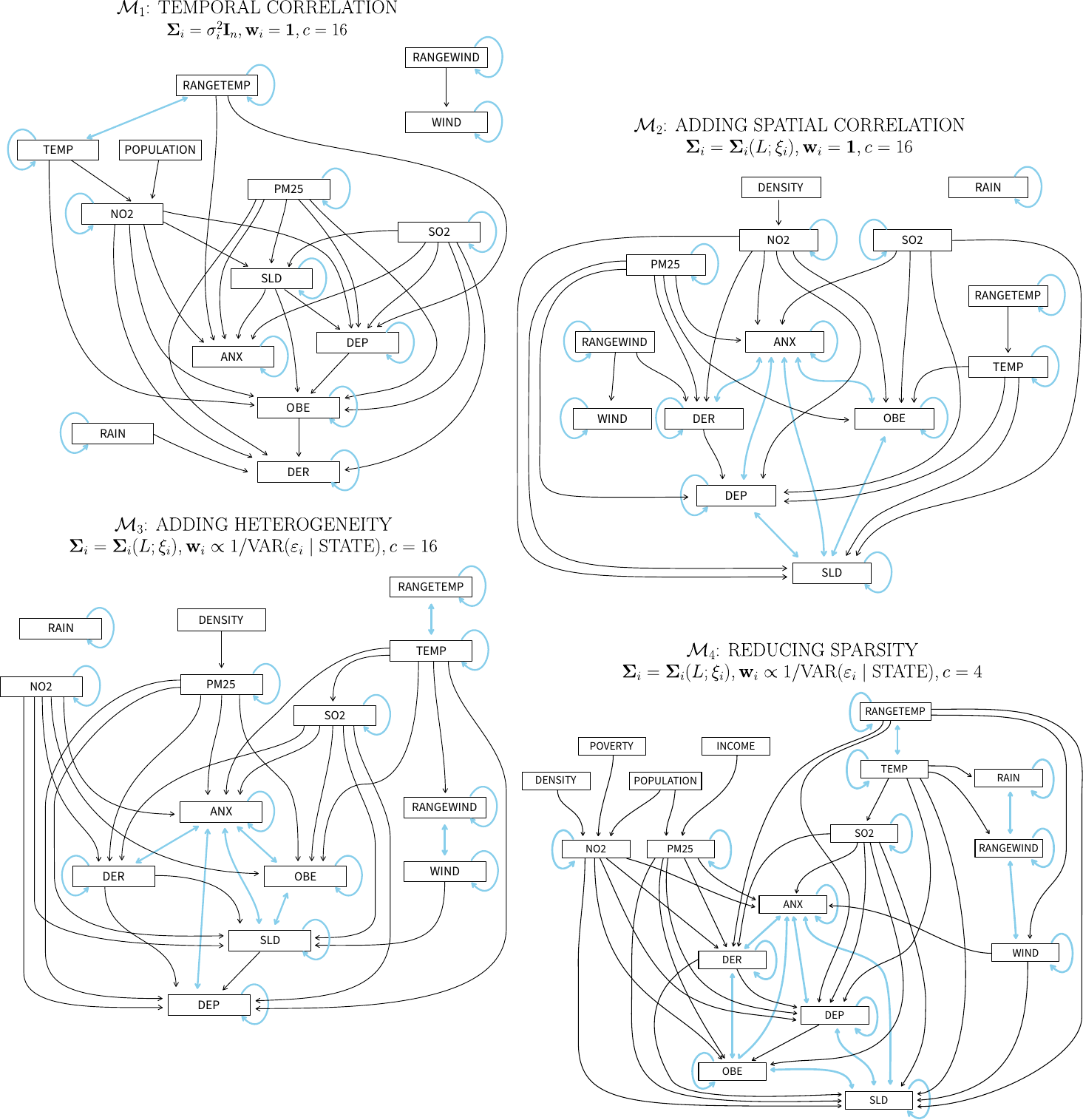}
  \caption{Biased CNs that only model part of the dependence structure between
    observations ($\M_1$, $\M_2$, $\M_3$) and our final CN ($\M_4$). Node
    labels: DER (dermatitis), OBE (obesity), ANX (anxiety), DEP (depression),
    SLD (sleep disorders), TEMP (temperature), RANGETEMP (thermal excursion),
    WIND (wind speed), RANGEWIND (wind speed range), RAIN (precipitation),
    POVERTY (poverty rate), DENSITY (population density), POPULATION (population
    size), INCOME (household income). Isolated nodes are not shown. Arcs in
    light blue represent feedback loops between two different variables or
    temporal autocorrelation for a single variable.}
  \label{fig:dbns}
\end{figure}

For the data described in Section~\ref{sec:data}, we assume an exponential
correlation structure for the $\S{i}(\L; \xxi{i})$ where $\xxi{i}$ are the range
and nugget estimated for each $X_{it}$. All $X_i$s have already been adjusted
for the counties' different areas and populations. They are aggregated over a
large population and a whole week: their distribution is approximately normal,
so we use classical GLS regressions and $\f{i}(\cdot) = I(\cdot)$. Google
normalised the search frequencies for anxiety, depression, sleep disorders,
obesity and dermatitis by state, so we set $\w{it} = 1/ \VAR(\err{it} \given
\mathrm{STATE})$ to adjust for that. To represent our fundamental understanding
of the causal relationships between the variables, we impose a topological
ordering that sets the demographic and weather variables at the top of the CN,
followed by the pollutants and then by the mental conditions, obesity and
dermatitis. We further blacklist arcs between the demographic variables, which
are static, and between pollutants, which cannot chemically change into each
other. We implement structure learning using model averaging (100 bootstrap
samples), tabu search and the PNAL score with penalties \mbox{$k = c \log(n)/2$,
$c = 2, 4, 8, 16, 32$} to model missing data and control sparsity.

\section{Results}
\label{sec:evaluation}

Figure \ref{fig:dbns} shows the CNs we evaluated while designing the model
specification in Section~\ref{sec:definition}: $\M_1$ accounts only for missing
data and temporal structure; $\M_2$ further models spatial correlation; $\M_3$
models the heterogeneity as well; and the final model $\M_4$ in which we relax
the sparsity penalty $c$ to let additional known causal relationships into the
CN.

Only $\M_4$ has a structure that captures the relationships we know exist from
the literature (Section~\ref{sec:data}), linking weather patterns, pollutants,
mental conditions and dermatitis. All models broadly identify pollutants as risk
factors for mental conditions \cite{buoli} and dermatitis \cite{park}. $\M_1$
misses the feedback loops between anxiety, sleep disorders and depression, and
those between obesity, sleep disorders and dermatitis we established in
\cite{loreal22}. $\M_2$ captures them to a greater extent but fails to link
weather with pollution. $\M_3$ captures this link and recognises the impact of
counties' socio-demographic characteristics on pollution. This is expected:
unlike $\M_1$ and $\M_2$, $\M_3$ is correctly specified for the data and can
better capture its underlying causal effects. However, the sparsity imposed on
$\M_3$ by choice of the PNAL penalty $c$ leads to discarding many smaller but
essential causal effects, such as that of temperature on dermatitis
\cite{engebretsen}. Reducing $c$ allows the inclusion of this link in $\M_4$.
The weather variables now form a connected component, whereas rain was
disconnected from the rest and from the pollutants in the other models.

We cannot rank these models as we did above using their predictive accuracy.
Even though they encode very different sets of causal effects, they all have an
average predictive $R^2$ of $0.742$ to $0.759$ over the conditions for new time
points (training set: until 2022/05/23; validation set: from 2022/12/26) and of
$0.715$ to $0.732$ for new locations (cross-validation over six geographical
areas, each comprising 15--20\% of the data, training and validation separated
by 110km). The Bayes factors between $\M_1$, $\M_2$, $\M_3$, $\M_4$ are
\mbox{$\BF_{12} = 95.27$}, \mbox{$\BF_{23} = 24.85$}, \mbox{$\BF_{34} = 2.90$}.
They would suggest that we are increasingly close to overfitting and that we should
choose one of the simpler, misspecified models.

The statistical tests from Section~\ref{sec:definition} correctly identify which
models are appropriate for the data. Simpler models are biased; their predictive
accuracy is inflated by bias rather than arising from accurate causal effect
estimates. $\M_3$ and $\M_4$ pass all checks: conditions have no unmodelled
temporal dependencies (0.1\% of p-values < 0.05 across lags 1--8), negligible
unmodelled spatial dependencies (2.2\% of p-values < 0.05), and no unmodelled
heterogeneity (no p-values < 0.05). In contrast, $\M_1$ models temporal
dependencies correctly (0.3\% of p-values < 0.05) but has substantial unmodelled
spatial dependencies (51.2\% of p-values < 0.05). $\M_2$ incorporates spatial
dependencies correctly (1.9\% of p-values < 0.05), but residuals are markedly
heterogeneous (all p-values < $1e^{-100}$).

Therefore, our final model $\M_4$ passes both a statistical validation (of its
assumptions and predictive accuracy over time and space) and domain evaluation
(of the causal effects it encodes) based on the literature. \\[-1.075\baselineskip]

We can formulate novel hypotheses from it by performing probabilistic and causal
queries on the effect sizes and mediation patterns of dermatitis risk
factors. Literature sources typically cannot answer such queries because they do
not simultaneously investigate different classes of risk factors, and the data
they use are limited. Experts are likewise limited by their focus on the
respective fields of specialisation, which may limit their knowledge of other
fields.

Firstly, we ask: \emph{what is the (simultaneous) relative impact of the direct
risk factors for dermatitis?} They are the parents of the corresponding node
(DER) in $\M_4$: thermal excursion (RANGETEMP), anxiety (ANX), obesity (OBE),
NO\sub{2}, SO\sub{2}, PM\sub{2.5}. After discounting autocorrelation, which
explains part of DER at a given time from its prevalence in the previous week,
the proportion of explained variance attributable to ANX is 0.574, followed by
NO\sub{2} (0.339), OBE (0.077), PM\sub{2.5} (0.008), RANGETEMP (0.001) and
SO\sub{2} (0.001).

We also ask: \emph{what proportion of the effects of environmental risk
factors is mediated by mental conditions?} $\M_4$ draws a complex network of
direct and indirect effects from the weather and pollution to dermatitis,
represented by arcs incident on DER and paths that step through other variables
before reaching it. Consider the pollutants: we can measure the proportion of
the variance of DER they explain with a lag of 1 month, block their effect on
mental conditions with a causal intervention, and then measure the remaining
(direct) effect on DER. The effects of PM\sub{2.5}, NO\sub{2} and SO\sub{2}
change by a factor of 0.54, 0.93 and 0.56. In absolute terms, however, the
overall effect of NO\sub{2} is about 11 times larger than that of PM\sub{2.5}
and 16 times larger than that of SO\sub{2}. Therefore, $\M_4$ suggests that
reducing SO\sub{2} has the greatest impact on dermatitis because it explains
much of its variability and its effect is only partly mediated by other
variables.

We can ask the same question for the weather variables: \emph{what proportion of
their effects is mediated by the pollutants and the mental conditions?} The
combined proportion of variance explained by TEMP and RANGETEMP changes by 0.11
(0.29) after removing the mediation effect of mental conditions (and pollution);
WIND and RANGEWIND change by 0.05 (0.38); and RAIN changes by 0.48 (0.02). In
proportion, the variance TEMP, RANGETEMP, WIND and RANGEWIND increases because
the effect of the mediating variables was in the opposite direction compared to
the respective direct effects. These effects are comparatively small,
suggesting that changing weather patterns will have a limited impact on
dermatitis.

These findings align with existing literature. NO\sub{2} exposure is documented
to correlate with increased dermatological outpatient visits more than
PM\sub{2.5} and SO\sub{2} \cite{zhang}. The indirect pathway from NO\sub{2} to
DER through ANX encodes the documented association between short-term NO\sub{2}
exposure and increased hospital admissions for anxiety disorders \cite{ma}.
Overall, NO\sub{2} acts through both direct pro-inflammatory mechanisms and
indirect anxiety-mediated pathways.

Furthermore, \emph{what would be the impact of tightening environmental
regulations on dermatitis?} The EPA reduced the legal limit on PM\sub{2.5} by
$25\%$ to $9 \mu{g}/m^3$ in 2024 to prevent an expected 4500 premature
deaths/year. A causal intervention on PM\sub{2.5} to reduce new emissions by
25\% at all locations in $\M_4$ for a year would reduce DER by 18\%, as would
strict enforcement of the new limit. Smaller commitments, such as reducing
PM\sub{2.5} emissions above $9 \mu{g}/m^3$ by 50\%, would reduce DER only by
5\%. On the other hand, reducing PM\sub{2.5} to $8 \mu{g}/m^3$ would reduce DER
by 21\%, suggesting that further tightening environmental regulations would have
a marked effect on the prevalence of dermatitis.

Finally, we can use counterfactuals to investigate how local differences in
space or time change the impact of environmental effects and how quickly they
propagate while effectively controlling for all other factors. For instance, we
can ask: \emph{how long must a cold spell last before we see an increase in
dermatitis?} This amounts to choosing a low starting temperature (<10$^\circ$C)
and a counterfactual higher temperature (> 20$^\circ$C) at a given time and
examining the difference in the distribution of DER at increasing time lags.
$\M_4$ finds a meaningful increase only after 4 weeks (+5\%).

\section{Conclusions}
\label{sec:discussion}

Using CNs to learn complex models of environmental and clinical risk factors in
infodemiology produces insights that are impossible to investigate using more
limited epidemiological data and comorbidity studies. We have demonstrated how
to do that for dermatitis and the complex interplay between weather, pollution
and psychological risk factors.

The approach we proposed is designed to produce CNs that correctly model the
spatial, temporal and incomplete data patterns typical of infodemiological data
arising from the fusion of different internet databases. We also provided
theoretical guarantees about the identifiability of causal effects, a
framework for statistical validation that goes beyond predictive
accuracy, and provided novel answers to several epidemiological questions.

Our CN has important implications for public health policy. Intervention
analysis emphasises that even modest adjustments in environmental regulations
can have a marked impact on health outcomes. Mental health and pollution are
individually strong risk factors: addressing both may generate synergistic
benefits. Mediation analysis confirms that pollution reduction policies may also
improve mental health, in turn reducing the prevalence of dermatitis
attributable to them. This highlights the value of integrated policies
addressing both environmental and mental health factors to maximise public
health benefits.

% \bibliographystyle{splncs04}
% \bibliography{references}

\end{document}